# Rapid Phase Retrieval by Lasing

Chene Tradonsky[1], Oren Raz[1], Vishwa Pal[2], Ronen Chriki[1], Asher A. Friesem[1] and Nir Davidson[1]*

**Reconstructing an object solely from its scattered intensity distribution is a common problem that occurs in many applications. Currently, there are no efficient direct methods to reconstruct the object, though in many cases, with some prior knowledge, iterative algorithms result in reasonable reconstructions. Unfortunately, even with advanced computational resources, these algorithms are highly time consuming. Here we present a novel rapid all-optical method based on a digital degenerate cavity laser, whose most probable lasing mode well approximates the object. We present experimental results showing the high speed (<100 ns) and efficiency of our method in agreement with our numerical simulations and analysis. The method is scalable, and can be applicable to any two dimensional object with known compact support, including complex-valued objects.**

Calculating the intensity distribution of light scattered far from a known object is relatively easy: it is the square of the absolute value of the object's Fourier transform[1]. However, reconstructing an object from its scattered intensity distribution is generally an ill-posed problem, because the phase information is lost and different choices of phase distributions result in different reconstructions. Fortunately, in many applications additional prior information, e.g. the object's shape, positivity, spatial symmetry, or sparsity, can be exploited to remove extraneous phase distributions, and hence only retrieve the original phase distribution and enable object reconstruction. Examples for applications can be found in astronomy[2], short pulses characterization[3], X-ray diffraction[4,5], radar detection[6], speech recognition[7] and imaging through turbid media[8,9].

For objects with a finite extent (compact support), a unique solution to the phase retrieval problem almost always exists (up to trivial ambiguities), provided that the scattered intensity is sampled at a sufficiently high resolution[10]. During the last decades, several algorithms for solving the phase retrieval problem have been developed. These include the Gerchberg-Saxton (GS) error reduction algorithm[11], hybrid input-output (HIO) algorithm[12], relaxed averaged alternating reflections (RAAR) algorithm[13], difference map algorithm[14], and shrink-wrap algorithm[15] (see Refs [16–18] for a modern review). Unfortunately, these algorithms are based on iterative projections and hence are relatively slow even with high performance computers[17,19].

Here we present and experimentally demonstrate a novel all-optical system that can solve phase retrieval problems rapidly. It is based on a digital degenerate cavity laser (DDCL)[20,21], into which two constraints - the Fourier magnitudes of the scattered light from an object and the compact support - are incorporated. Then, the nonlinear lasing process results in a self-consistent solution that best satisfies both constraints. An upper bound of 100ns was measured for the time needed by the DDCL to converge to a stable solution.

In several computational challenges, specifically tailored physical systems could be more efficient than conventional silicon based computers. These systems are not universal Turing machine, namely they cannot perform any calculation as a standard computer, but they can solve a specific class of problems very efficiently. Prominent examples for such systems are the D-Wave machine that searches for the ground state of a complicated Hamiltonian through quantum annealing[22–24], and coupled lasers, optical parametric oscillators (OPO) and coupled polaritons systems that can solve various difficult optimization problems[25–32]. Solving hard problems with such systems offer significant advantages in computation time and resources over conventional computers[25,33].

The physical mechanism that generates the nonlinear lasing process in DDCLs is similar to that of the OPO spin simulators[27,28,30,34]. Both the OPO simulators and the DDCLs have distinct advantages in performing optimization, including extremely fast operation[25,30], ability to avoid local minima when the pumping is increased slowly enough[27,29] and having a non-Gaussian wave-packet[35]. Our DDCL system has several particular attractive and important features. These include high parallelism that provides thousands of parallel realizations simultaneously – one in each coherence length of the amplified spontaneous emission (ASE) before lasing; short round-trip times (~20ns), leading to fast convergence time; and inherent selection of the mode with minimal loss (optimal solution to a particular problem), due to mode competition.

**Experimental arrangement**

The basic DDCL arrangement for rapidly solving the phase retrieval problem is schematically presented in Figure 1. It consists of a ring degenerate cavity laser that includes a gain medium, two 4f telescopes, an amplitude spatial light modulator (SLM), an intra-cavity aperture, three high reflectivity mirrors and an output coupler. The left 4f telescope ($f_1$ and $f_2$) images the center of the gain medium onto the SLM, where the transmittance at each pixel is controlled independently[20]. The intra-cavity aperture, together with the SLM, serve to control and form the output lasing intensity distribution[36,37].

In the absence of the intra-cavity aperture, the right 4f telescope simply reimages the SLM back onto the gain medium, so all phase distributions can lase with equal probability, i.e. the amplification and losses are phase independent. However, when an intra-cavity aperture (compact support mask) is placed at the Fourier plane between the two lenses, each phase distributions has a different level of loss. In this case, the phase distribution that experiences the minimal loss is the most probable lasing mode, due to mode competition. With the Fourier intensity distribution of the original object applied onto the SLM and the appropriate compact support imposed by the intra-cavity aperture, (the two constraints of the phase retrieval problem), the most probable lasing mode corresponds to the optimal solution of the phase retrieval problem. The field distributions of the other lasing modes spread beyond the boundaries of the intra-cavity aperture, and therefore suffer from loss. The field distribution of the most probable lasing mode, i.e. of the reconstructed object, is formed within the intra-cavity aperture and is imaged through the output coupler onto the camera.

[1]Department of Physics of Complex Systems, Weizmann Institute of Science, Rehovot 7610001, Israel. [2]Department of Physics, Indian Institute of Technology Ropar, Ropar 140001, India.
*e-mail: nir.davidson@weizmann.ac.il

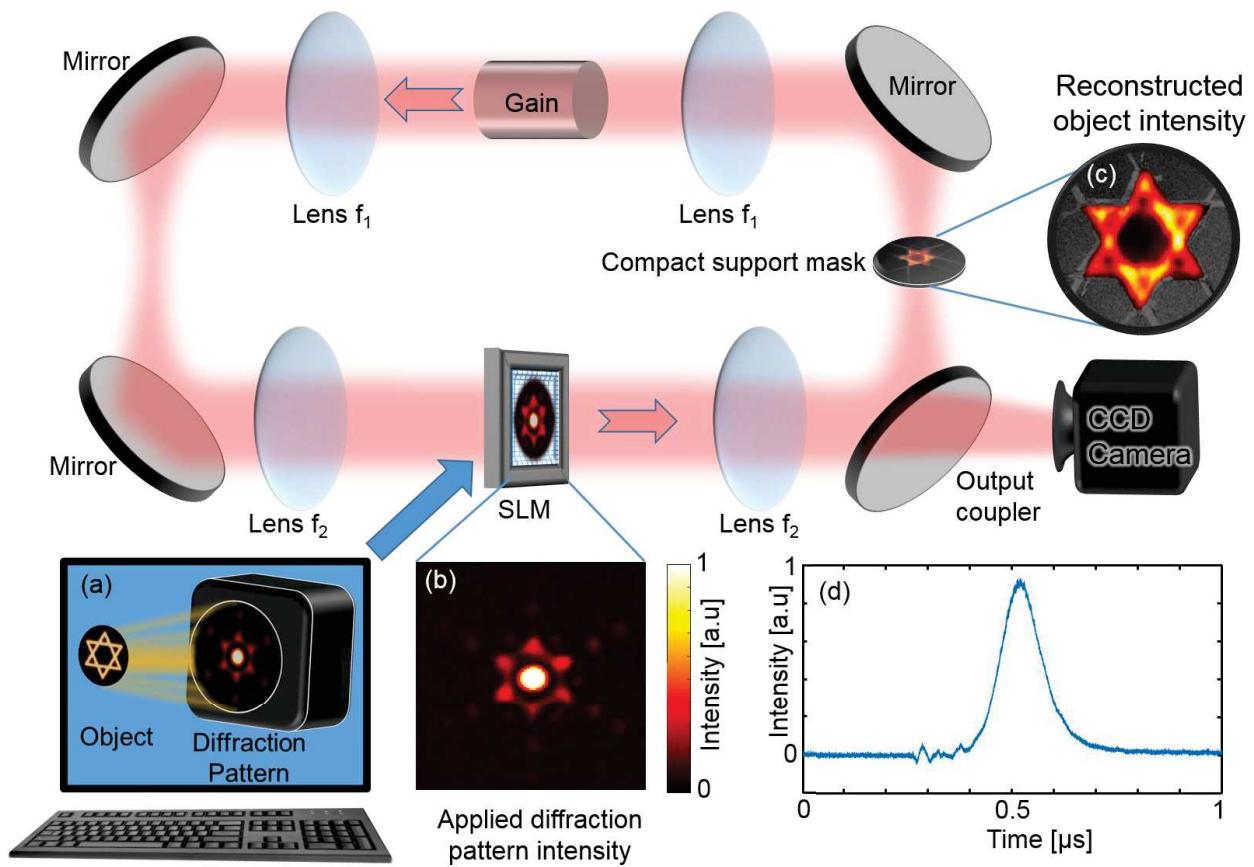

**Figure 1. Basic digital degenerate cavity laser arrangement for rapid phase retrieval.** (a) Calculated scattered intensity distribution from the object is applied onto a spatial light modulator (SLM), which is incorporated into a ring degenerate cavity laser that can support up to 100,000 degenerate transverse modes. A mask shaped as the object boundaries (compact support) at the Fourier plane filters out extraneous modes that do not match the compact support. With this laser arrangement, the lasing process yields a self-consistent solution that satisfies both the scattered intensity distribution shown in (b) and the compact support constraint. (c) The reconstructed object intensity appears at the compact support mask and imaged onto the camera. (d) Laser intensity as a function of time. The duration of the lasing process (convergence to a solution) is limited to about 100ns by incorporating a Pockels cell into the laser cavity (see Figure S2).

**Experimental results.**

Representative experimental results for different objects are presented in Figures 2-4. Columns (a) show the actual objects, columns (b) show the corresponding Fourier intensity distributions, which were used to control the transmission of the SLM (see supplementary), and columns (c) [and column (d) in Figure 4] show the detected intensity distributions of the reconstructed objects along with the compact support outlines.

Figure 2 shows results for three centrosymmetric objects with uniform phase distribution (i.e. real valued objects) and circular compact support. As evident, there is very good agreement between the intensity distributions of the actual objects and those of the reconstructed objects.

Figure 3 shows the results for objects with centrosymmetric intensity distributions and various complex phase distributions, and circular compact support. The first row shows the results for an object with uniform phase distribution, where the corresponding scattered intensity distribution has a 12-fold symmetry. As evident, the reconstructed object is very similar to the actual object. The second row shows an object with centrosymmetric phase distribution, so both the object and the corresponding Fourier intensity distribution are centrosymmetric. Note that our system correctly reconstructs the actual object in spite of the strikingly different Fourier intensity distributions for rows one and two. The high quality reconstruction verifies that our approach is also valid for complex-valued objects, which are generally harder to solve computationally[38].

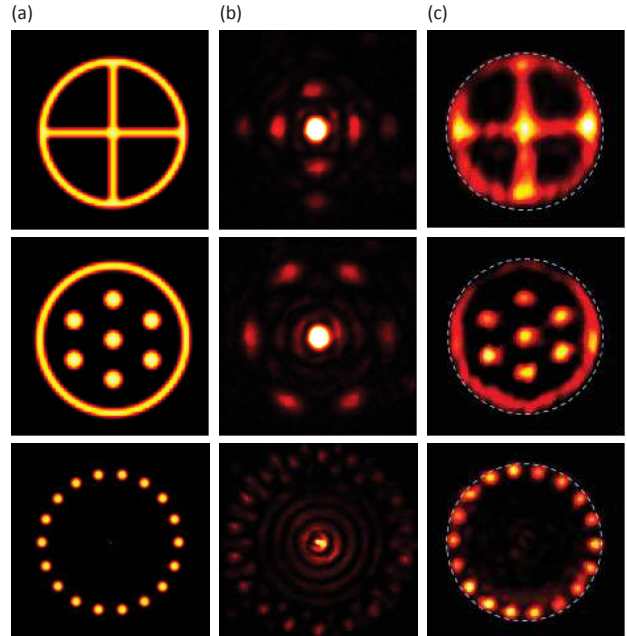

**Figure 2. Experimental results for real valued centrosymmetric objects.** Column (a) - intensity distributions of the actual objects. Column (b) - corresponding Fourier intensity distributions, applied to control the SLM. Column (c) - detected intensity distribution of the reconstructed objects, using a circular aperture as compact support.

A known ambiguity in phase retrieval emerges when the object is non-centrosymmetric, but the assumed compact support is centrosymmetric. Example for such a case is shown in the third row of Figure 3, where the object has a random, asymmetric phase distribution, so the corresponding Fourier intensity distribution is also asymmetric. As evident, the blurred reconstructed object differs from the actual object due to interferences between two degenerate solutions (one of the image of the object and the other of the inverted phase conjugated image). Applying non-centrosymmetric compact support (for a non-centrosymmetric object), removes this degeneracy, and ensures high quality reconstruction, as seen in the fourth row of Figure 3.

actual object distribution is compatible with the constraints. Note that a similar approach is commonly used in several phase retrieval algorithms in order to resolve ambiguities and improve the reconstructed object[39].

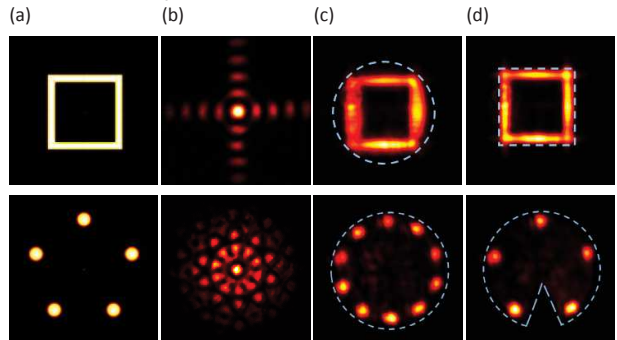

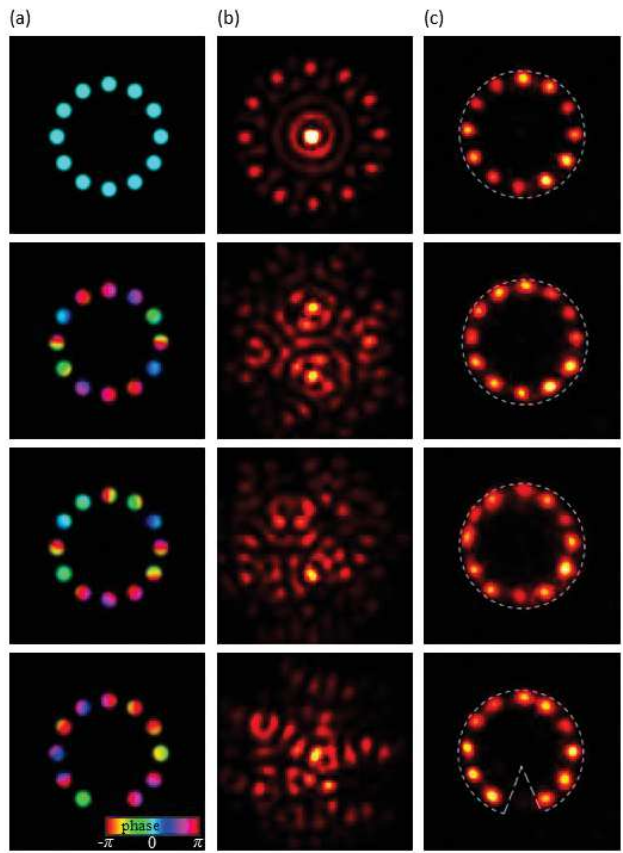

**Figure 3. Experimental results for complex-valued objects.** Column (a) - intensity (brightness) and phase (hue) distributions of the actual objects. Column (b) - corresponding Fourier intensity distributions, applied to control the SLM. Column (c) - detected intensity distribution of the reconstructed objects, using mainly a circular aperture as compact support. The first row shows an object with uniform phase distribution; the second [third] row shows the same object with arbitrary centrosymmetric [asymmetric] phase distribution; the fourth row shows a non-centrosymmetric object with random asymmetric phase distribution and non-circular compact support.

We also investigated the effect of tightness and symmetry of the compact support on the reconstruction quality. Representative experimental results are presented in Figure 4. The results in the first row demonstrate that a tight compact support (square rather than a circular aperture) significantly improves the quality of reconstructed square object. Yet, since both the object and the support are centrosymmetric, even the non-tight circular aperture leads to a reasonable reconstruction. The results in the second row demonstrate the importance of centrosymmetry. The object in this case is non-centrosymmetric, so when the compact support is centrosymmetric (circular aperture in column c), two different solutions - the reconstructed object and its centro-inverted version - are compatible with the constraints[10], and as evident both are obtained. However, by adding a wedge to the compact support, the centrosymmetry of the compact support is broken, and only the

**Figure 4. Experimental results demonstrating the effect of tightness and asymmetry of compact supports.** Column (a) - intensity distribution of the actual objects. Column (b) - corresponding Fourier intensity distributions, applied to control the SLM. Column (c) detected intensity distribution of the reconstructed objects, using a circular aperture as compact support. Column (d) - detected intensity distribution of the reconstructed objects, using a square aperture as tight compact support (upper row) and a circular aperture with a wedge as asymmetric compact support (lower row).

Typically, the resolution of the reconstructed objects was relatively low (about 20x20). We attribute the low-resolution mainly to phase aberrations in our laser cavity. Correcting these, e.g. with our intra-cavity SLM, can significantly improve the resolution, as indicated by our numerical simulations (see supplementary).

To determine an upper bound on the minimum duration needed for the laser to find the optimal lasing mode and reconstruct the object, we resorted to a Q-switched linear degenerate cavity laser arrangement that included a Pockels cell, as shown in Figure 2S. The shortest pulse we could generate was 100ns long (see pulse profile in Figure 1(d)), and even then the laser reconstructed the object successfully. In general, the results obtained with Q-switching were essentially the same as those with quasi-CW lasing operation. This indicates that the computation time of the system is at most 100ns. See supplementary for additional details.

**Discussion**

Let us now explain why the lasing mode of the DDCL system in Figure 1 corresponds to the solution of the phase retrieval problem. The lasing mode in the system is a complex field at the SLM, $E(\vec{k}, t)$ where $\vec{k}$ is the position at the SLM plane, mapped onto itself after propagating through the cavity. In other words, it is a stationary solution of the field propagation equation, which for a cavity round-trip of duration $\tau$ can be written as

$$E(\vec{k}, t+\tau) = T_{SLM}(\vec{k}) G(\vec{k}, t) \mathcal{F}^{-1}\left[M(\vec{x}) \mathcal{F}[E(\vec{k}, t)]\right], \quad (1)$$

where $T_{SLM}(\vec{k})$ is a linear transformation that represents the amplitude transmittances at the SLM, $G(\vec{k}, t)$ is the (nonlinear) gain of the system, $\mathcal{F}$ is 2D Fourier transform (performed by the lenses), and $M(\vec{x})$ is a linear transformation (a projection) that represents the spatial compact support imposed by the intra-cavity mask, where $\vec{x}$ is the position at the mask plane. Note, that the mapping of $E(\vec{k}, t)$ is nonlinear, due to the nonlinear gain with saturation[40] $G(\vec{k}, t) = g_0\left(1 + |E(\vec{k}, t)|^2/I_{sat}\right)^{-1}$, where $g_0$ is the linear gain at very low intensities set by the pumping strength, and $I_{sat}$ is the saturation intensity.

Now consider the electric field $E_{sol}(\vec{k})$, which corresponds to the solution for the phase retrieval problem. This field passes through the compact support without any changes, $\mathcal{F}^{-1}\left[M(\vec{x})\mathcal{F}[E_{sol}(\vec{k})]\right] = E_{sol}(\vec{k})$. Assuming that $E_{sol}(\vec{k})$ is a stable, time independent solution of Equation (1), so $T_{SLM}(\vec{k})G(\vec{k},t) = 1$, and the SLM transmittance at each pixel must be

$$T_{SLM}(\vec{k}) = \frac{1+|E_{sol}(\vec{k})|^2/I_{sat}}{g_0}. \quad (2)$$

With this choice for $T_{SLM}(\vec{k})$, the solution of the phase retrieval is a possible lasing mode in the system. It is important to note we do not need to resort to the unknown field $E_{sol}(\vec{k})$ in order to calculate $T_{SLM}(\vec{k})$, but only to its known scattered intensity distribution $|E_{sol}(\vec{k})|^2$. In other words, we can tune the system to the specific problem for which we wish to solve the phase by tuning $T_{SLM}(\vec{k})$ according to Equation (2).

Substituting the expressions for $T_{SLM}(\vec{k})$ in Equation (2) and $G(\vec{k},t)$ into Equation (1), yields

$$E(\vec{k}, t+\tau) = \frac{1+|E_{sol}(\vec{k})|^2/I_{sat}}{1+|E(\vec{k},t)|^2/I_{sat}} \mathcal{F}^{-1}\left[M(\vec{x})\mathcal{F}[E(\vec{k},t)]\right]. \quad (3)$$

Equation (3) can be considered as a modified GS iterative projection process, in which the fastest growing mode corresponds to the solution. To verify our approach, we performed numerical simulations of the cavity. We assumed that the initial state $E(\vec{k},0)$ is a random complex field. The simulation results, which have a similar behavior to the experimental results, are presented in Figure S3 in the supplementary. Additional details on the numerical simulations are given in the supplementary.

So far, we assumed that the solution for the phase retrieval problem is the only stable lasing mode when the SLM is properly tuned. However, there might be other stable lasing modes in our nonlinear cavity, which can lead to a wrong solution. Before the transition to lasing, the gain operates in the incoherent amplified spontaneous emission (ASE) regime, where the phases can be considered independent after each coherence length. Light from each coherence length of the ASE is initiated at a different random phase realization. As it propagates through the cavity, it evolves according to Equation (3). Since the cavity length is much longer than the coherence length of the ASE, we can view the ASE growing stage as a large number of parallel realizations, each of which evolves independently under the iterative projection process of Equation (3). For Nd-YAG around 1064nm, the ASE coherence length is about 2mm, hence in our 5m cavity there are about 2500 independent realizations is in a round-trip.

At the transition to lasing, the ASE modes with the highest energy win the mode competition over the limited gain. In the initial growth stage of the electric field inside the cavity, $|E(\vec{k},t)|^2$ of each ASE mode is extremely small. Therefore, Equation (3) can be approximated as

$$E(\vec{k}, t+\tau) \sim \left(1+\frac{|E_{sol}(\vec{k})|^2}{I_{sat}}\right)\mathcal{F}^{-1}\left[M(\vec{x})\mathcal{F}[E(\vec{k},t)]\right]. \quad (4)$$

Under this approximation, the round-trip mapping is linear. The fastest growing mode in this stage is hence the eigenmode of the linear mapping with the highest eigenvalue. When $I_{sat} \gg |E_{sol}(\vec{k})|^2$, Equation (4) is to a good approximation a projection on the compact support. Hence, all modes within the support grow exponentially faster than other modes. Thus, the solutions of the phase retrieval problem are both the fastest growing modes in the initial stage and the stable lasing modes. Moreover, since the phase retrieval problem has a unique solution, it assures that $E_{sol}(\vec{k})$ would be the only stable lasing mode from all the modes with a certain compact support. Thus, it is expected to be the most probable lasing mode.

## Concluding remarks

We presented an all-optical system for rapid phase retrieval, using a novel digitally controlled degenerate cavity laser (DDCL). A measured upper bound on the time needed to reconstruct an object just from its scattered intensity distribution was 100ns, orders of magnitude faster than conventional computation systems. Although the DDCL can solve phase retrieval problems in less than 100ns, setting the scattered intensity distribution as the input on the SLM could take at least few milliseconds. A direct approach, which uses the scattered light from the unknown object as on-axis structured pump could significantly speed up the process.

Several modifications to the system can potentially improve the performance. For example, resort to other numerical algorithms, such as hybrid input-output (HIO) algorithm[12] where a small feedback (e.g. implemented in our system with a delay line) can dramatically improve the rate of convergence of the system. Another example is to resort to a sparsity constraint, by means of a saturable absorber inside the cavity.

Finally, we believe that in addition to finding solutions to phase retrieval problems rapidly, our DDCL systems can be exploited for solving many other problems that occur in various fields, including three-dimensional object reconstruction and resolving imagery after propagation through scattering media[9].

## Methods

In our experiments, we actually used a reflective phase-only SLM, rather than the transmissive SLM in the laser arrangement shown in Figure 1. The reflective SLM has a relatively high light efficiency and high damage threshold. Accordingly, the laser arrangement was modified to retain the same operation functionality. The detailed experimental arrangement that includes the reflective SLM is schematically presented in Figure 1S in the Supplementary Material, along with an explanation on how a phase-only SLM together with the intra-cavity aperture can control the amplitude transmittance of each effective pixel in the SLM[27].

In our experimental arrangement, the laser gain medium was a 1.1% doped Nd-YAG rod of 10mm diameter and 11cm long. For quasi-CW operation, the gain medium was pumped high above threshold by a 100μsec pulsed Xenon flash lamp operating at 1500V and a repetition rate of 1 Hz to avoid thermal lensing. Each 4f telescope consists of two plano-convex lenses, with diameters of 50.8mm and focal lengths of $f_1 = 750$mm and $f_2 = 500$mm at the lasing wavelength of 1064nm. The SLM was Hamamatsu (LCOS-SLM X13138-03) with high reflectivity of about 98% at 1064nm wavelength, high resolution, and high damage threshold. For Q-switch operation, a Pockels cell was incorporated into a linear DCL of the same gain and pump as the quasi-CW operation, and the focal lengths of two lenses in the telescope were 250mm. Additional details are presented in the supplementary.

## Acknowledgements
This work was supported by the Israel Science Foundation, and the United States-Israel Binational Science Foundation. O.R. is supported by a research grant from the Center for New Scientists at WIS.

## Author contributions
All authors contributed to writing the manuscript. C.T. designed the experimental arrangement, performed the experiments and the numerical simulations with contributions and inputs from all authors. O.R. developed the theory.

## Competing interests
The authors declare no competing interests.

## Additional information
Supplementary information is available for this paper.

# Rapid Phase Retrieval by Lasing - Supplementary

Chene Tradonsky[1], Oren Raz[1], Vishwa Pal[2], Ronen Chriki[1], Asher A. Friesem[1] and Nir Davidson[1]*

**Detailed experimental arrangement.**

The detailed experimental arrangement of the digital degenerate cavity laser (DDCL) is schematically presented in Figure 1S. It consists of a ring degenerate cavity laser that includes a gain medium, two 4f telescopes with one common lens, a reflective phase only spatial light modulator (SLM), an intra-cavity aperture, two retroreflectors and pentaprism-like 90° reflector (all from high reflectivity mirrors), two polarizing beam splitters (PBS), two half-wave plates (λ/2) and a Faraday rotator.

The operation of the detailed arrangement is essentially the same as the basic arrangement presented in Figure 1. Each of the two 4f telescopes has one lens $f_1$ and a common lens $f_2$. The first telescope images the field distribution at the center of the gain medium onto the SLM where the reflectivity of each effective pixel[1] is controlled. The second telescope, which contains an intra-cavity aperture, images the field distribution at the SLM that will result in the lowest losses back onto the gain medium. Such a distribution is determined by the size and shape of intra-cavity aperture (compact support).

Since our SLM operates on axis and by reflection on horizontal polarized light, half of the ring degenerate cavity was designed as a twisted-mode[2] linear degenerate cavity[3] and the other half as regular ring cavity laser[3]. The two halves are connected by $PBS_1$, which separates the two counter propagating beams to two different cross-polarized paths. A large aperture Faraday rotator together with a half-wave plate (HWP) at 22.5° and another $PBS_2$ (which also serves as ~5% output coupler) enforce unidirectional operation of the ring cavity. A 90° reflector flips left and right areas of the beam. The left retroreflector can compensate for free propagation diffraction in the cavity. The right retroreflector can compensate for phase spherical aberrations in the cavity. A second HWP at 45° rotates the polarization from vertical to horizontal to pass through $PBS_1$.

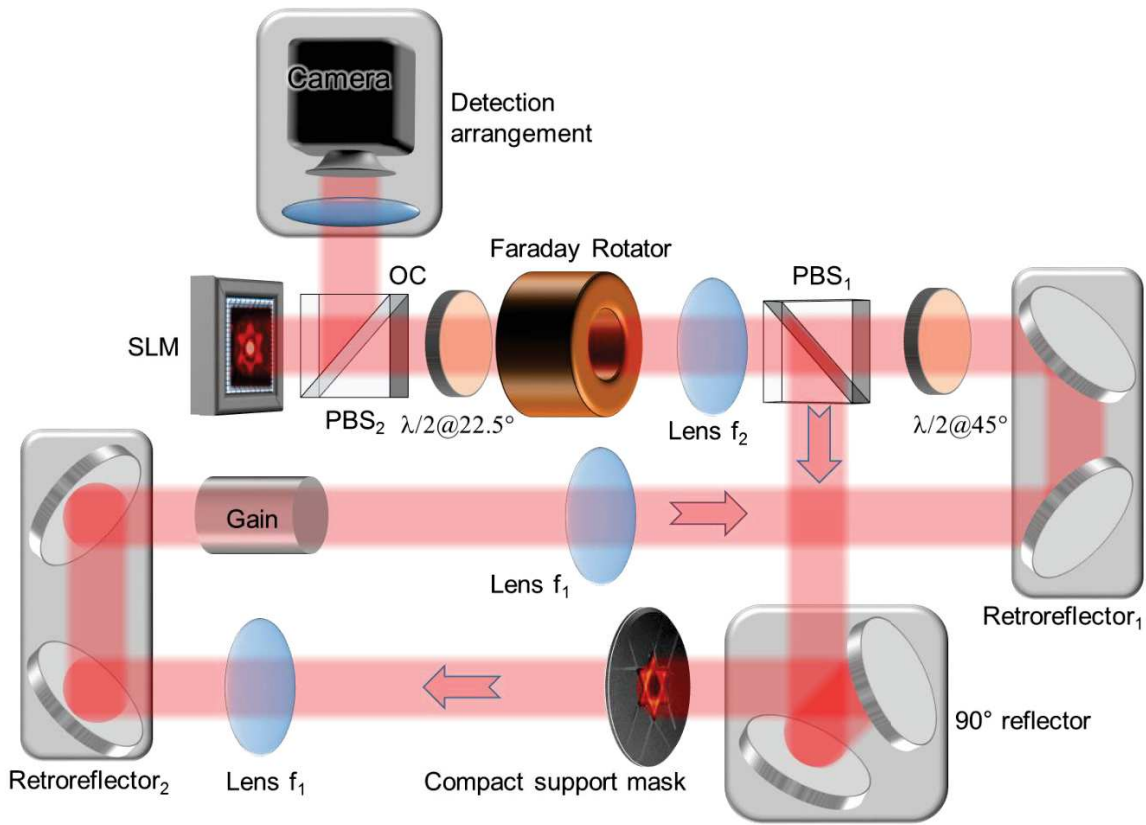

**Figure 1S. Detailed experimental digital ring degenerate cavity laser arrangement.** SLM - spatial light modulator; PBS - polarizing beam splitter; λ/2@22.5° - half-wave plate at 22.5° angular orientation; ; λ/2@45° - half-wave plate at 45° angular orientation; compact support mask - intra-cavity aperture at the Fourier plane; OC – output coupler.

[1]Department of Physics of Complex Systems, Weizmann Institute of Science, Rehovot 7610001, Israel. [2]Department of Physics, Indian Institute of Technology Ropar, Ropar 140001, India.
*e-mail: nir.davidson@weizmann.ac.il

The detection arrangement includes a CMOS camera and lenses so both the reconstructed object and the scattered intensity distributions can be detected.

The local reflectivities of the SLM are determined by the local phase difference of adjacent pixels that affects the amount of light diffracted outside the cavity, and the phases are determined by the local phase average of the adjacent pixels[1]. For example, adjacent pixels with phases of [0, 0] will result in high reflectivity and 0 phase, whereas adjacent pixels with phases of [0, π] will result in no reflectivity and π/2 phase. The reflectivity pattern can be used to form any desired intensity distribution and the phase distribution can be used to overcome aberrations in the cavity to increase the laser degeneracy. For the phase retrieval problem, the SLM was controlled such that its reflectivity pattern matches the intensity distribution, but does not add any relative phases between the pixels. Therefore, the lasing frequencies of all the pixels are identical, leading to an interference pattern in the Fourier plane (i.e. in the compact support mask location), which is the solution to the phase retrieval problem (reconstructed object).

**Convergence time to reach a solution**

In order to determine the time it takes the laser to solve the phase retrieval problem, we limited the lasing duration to a narrow pulse. For this purpose, we resorted to a Q-switched linear DCL, schematically presented in Figure 2S (a). The Q-switched linear DCL consists of a two lenses in a 4f telescope with a Pockels cell and two intra-cavity amplitude masks. The first mask was placed near the rear mirror and enforced a specific scattered intensity distribution (according to the phase retrieval problem). Here we used a metallic binary amplitude mask instead of a SLM, due to high peak power of Q-switched operation of the laser and the limited damage threshold of the SLM. The second mask was placed between the two lenses and served as the compact support of the reconstructed object.

The results are presented in Figures 2S (b-e). These show the intensity distributions at the mask (representing the scattered intensity distribution from the unknown object) and the intensity distributions of the reconstructed object at the compact support plane. Figures 2S (b) and (c) show the results at quasi-CW lasing (no Q-switching), and Figures 2S (d) and (e) show the results at Q-switched lasing operation with pulse duration of 100ns (shown in Figure 1 (d)). As evident, the short duration of the pulse does not affect the quality of the reconstructed object. Although we estimate the convergence time to reach a solution would be significantly shorter than 100ns, we set this value as the upper bound

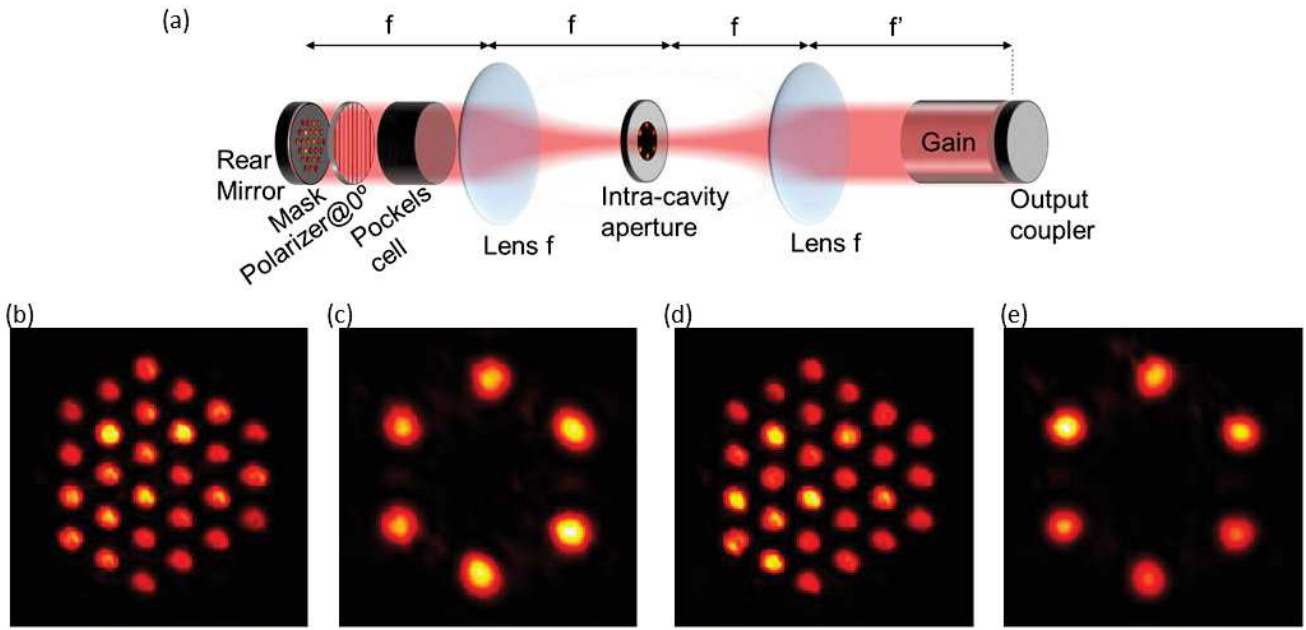

**Figure 2S. Experimental arrangement and results demonstrating rapid solutions**. (a) A linear Q-switched DCL with Pockels cell and two intra-cavity masks. (b) and (c) Scattered and reconstructed object intensity distributions, at quasi-continuous lasing operations (without Q switch). (d) and (e) Scattered and reconstructed object intensity distributions, at Q-switched lasing operations with pulse duration of 100ns.

**Simulation results.**

In order to support our experimental results, we performed some basic simulations. Specifically, we simulated the field distribution of the transverse mode inside the laser cavity (representing the reconstructed object) in the arrangement shown in Figure 1, by resorting to a modified Gerchberg-Saxton (GS) iterative algorithm[4]. In the simulation, we included the phase only SLM, laser gain medium and the aperture shaped as compact support of the object. We started with an initial guess of a random field distribution at the SLM plane, and then resorted to the iterative algorithm according to Equation (3). In each iteration, the field of the next round trip was calculated from the current one.

Representative simulation results are presented in Figure 3S. Figure 3S (a) shows an image of the actual scattering object. Figure 3S (b) shows the simulated intensity distribution of the diffraction pattern inside the cavity. Figure 3S(c) shows the reconstructed intensity distribution of the object after 100 iterations inside the laser cavity. The compact support shape in this example was the outer boundary, while the details inside were reconstructed by the algorithm. Note the simulation only deal with a single realization, out of the thousands parallel realizations that run in the cavity and compete on the gain.

We also investigated the effect of phase aberrations, spherical and others that are caused by the SLM, on the resolution of the reconstructed object. This was done by incorporating phase aberrations into the modified GS algorithm simulations. The results are presented in Figure 4S. Figures 4S (a) (b) and (c) show the simulation results for a low-resolution object and without any aberrations in our system. Figure 4S (a) shows the intensity distribution of the actual object, Figure 4S (b) the corresponding simulated intensity distribution of the diffraction pattern inside the cavity and Figure 4S (c) the intensity distribution of the reconstructed object after 100 iterations inside the laser cavity.

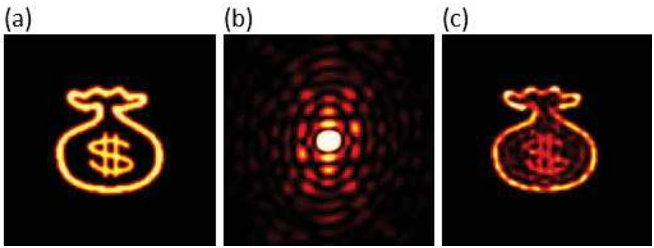

**Figure 3S. Representative simulation results of inverse problem solutions with a digital degenerate cavity laser.** (a) Image of the actual scattering object. (b) Simulated intensity distribution of the diffraction pattern inside the cavity and (c) intensity distribution of the reconstructed object after 100 iterations inside the cavity.

Figures 4S (d) (e) and (f) show the simulation results after adding typical aberrations from the SLM manufacturer calibration file (contains phase corrections to make SLM surface flat). Figure 4S (d) shows the typical phase aberrations distribution caused by the SLM, Figure 4S (e) the corresponding simulated intensity distribution of the diffraction pattern inside the cavity, and Figure 4S (f) the intensity distribution of the reconstructed object after 200 iterations inside the laser cavity.

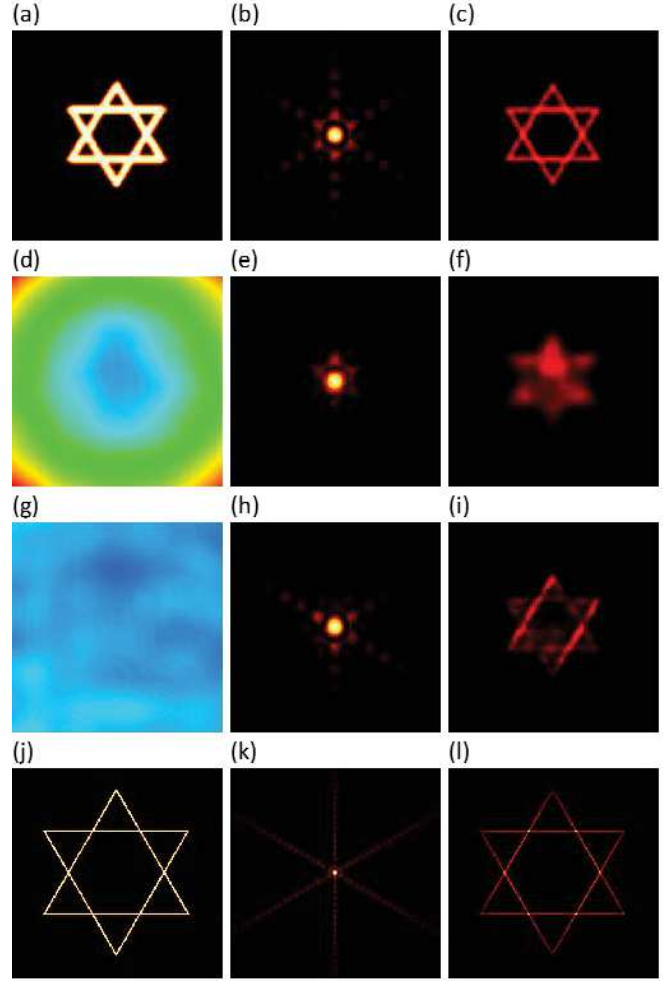

**Figure 4S. Representative simulation results with phase aberrations.** (a) Image of the actual scattering object at low-resolution (LR). (b) Simulated intensity of the LR diffraction pattern inside the cavity and (c) intensity distribution of the reconstructed object after 100 iterations inside the cavity. (d) Typical phase aberration of SLM (from the calibration file of the SLM). (g) Typical phase aberration of SLM after subtraction of spherical aberration. (e) and (h) Simulated intensity of the LR diffraction pattern inside the cavity and (f) and (i) intensity distribution of the reconstructed object after 200 iterations inside the cavity with phase aberration in (d) and (g). (j) Image of the actual scattering object at high-resolution (HR). (k) Simulated intensity of the HR diffraction pattern inside the cavity and (l) intensity distribution of the reconstructed object after 200 iterations inside the cavity.

Figures 4S (g) (h) and (i) show the simulation results after subtracting spherical aberrations but retaining the others. Figure 4S(g) shows the improved phase aberrations distribution, Figure 4S(h) the corresponding simulated intensity distribution of the diffraction pattern inside the cavity and Figure 4S (i) the intensity distribution of the reconstructed object after 200 iterations inside the laser cavity.

These results indicate that phase aberrations strongly affect the ability of the system to successfully reach the correct solution and reconstructed object that accurately matches the actual object. For example, Figure 4S (f) shows a poor reconstructed object when all phase aberrations of the SLM are taken into account. There is improvement when the spherical aberrations are removed, as shown in Figure 4S (i). This correction of the phase aberrations is done using a linear shift (right or left) of the retroreflector 1 in Figure 1S. Then the resolution of the reconstructed object is similar to that of the experimental results in Figures 2-4.

In order to ensure that our approach can potentially lead to perfect reconstructed objects, we performed simulations using high-resolution object and the assumption of no phase aberrations (perfect cavity). The results are shown in Figures 4S (j) (k) and (l). Figure 4S (j) shows the intensity distribution of the actual object, Figure 4S (k) the simulated intensity distribution of the diffraction pattern inside the cavity, and Figure 4S (l) the intensity distribution of the high-resolution reconstructed object after 200 iterations inside the laser cavity. This result clearly verifies the efficacy of our approach.